\documentclass[12pt,preprint]{aastex}

\shortauthors{Harker et al.}
\shorttitle{Disks of Three Herbig Stars}

\begin{document}

\title{The Disk Atmospheres of Three Herbig Ae/Be Stars}

\author{David E.\ Harker \altaffilmark{1}}
\affil{Center for Astrophysics and Space Sciences, University of California, 
San Diego, 9500 Gilman Drive, La Jolla, CA 92093-0424}

\author{Charles E.\ Woodward \altaffilmark{1}} \affil{Astronomy Department, 
School of Physics and Astronomy, 116 Church Street, S.E., University of 
Minnesota, Minneapolis, MN 55455} 

\author{Diane H.\ Wooden \altaffilmark{1}, Pasquale Temi}
\affil{NASA Ames Research Center, Space Science Division, MS245-1, Moffet 
Field, CA 94035-1000} 

\altaffiltext{1}{Visiting Astronomer, Cerro Tololo Interamerican 
Observatory, National Optical Astronomy Observatories, operated by the 
Association of Universities for Research in Astronomy, Inc., under 
cooperative agreement with the National Science Foundation.}


\begin{abstract}

We present infrared (IR) spectrophotometry ($R \simeq 180$) of three 
Herbig Ae/Be stars surrounded by possible protoplanetary disks: HD~150193, 
HD100546 and HD~179218.  We construct a mid-IR spectral energy 
distributions (SED) for each object by using $7.6 - 13.2$~\micron\ HIFOGS 
spectra, 2.4 -- 45~\micron\ spectrophotometry from the {\it ISO} SWS, the 
12, 25, 60, and 100~\micron\ photometric points from IRAS, and for 
HD~179218, photometric bolometric data points from the Mt.\ Lemmon 
Observing Facility.  The SEDs are modeled by using an expanded version of 
the \citet{chigol97} two-layer, radiative and hydrostatic equilibrium, passive 
disk.  This expanded version includes the emission from Mg-pure crystalline 
olivine (forsterite) grains in the disk surface layer.  Each of the three 
objects studied vary in the amount of crystals evident from their 
spectrophotometry.  HD~150193 contains no crystals while HD~100546 and 
HD~179218 respectively show evidence of having crystalline silicates 
in the surface layers of their disks.  We find that the inner region of 
HD100546 has a 37\% higher crystalline-to-amorphous silicate ratio in its 
inner disk region ($\leq 5$~AU) compared to the outer disk region, while the 
inner disk region of HD 179218 has a 84\% higher crystalline-to-amorphous
silicate ratio in its inner disk region ($\leq 5$~AU) compared to the outer 
region.  All three objects are best-fit using a grain size distribution power 
law which falls as $a^{-3.5}$.  HD~150193 is best-fit by a small disk 
($\sim 5$~AU in radius) while HD~100546 and HD~179218 are best fit by larger 
disks ($\sim 150$~AU in radius).  Furthermore, HD~100546's disk flares larger
than compared to HD~150193 (25\% more at 5~AU) and HD~179218 (80\% more at
5~AU).  We discuss the implications of our results and compare them with 
other modeling efforts. 

\end{abstract}


\section{INTRODUCTION}
\label{intro}

The ubiquity of accretion disks around pre-main sequence and young main 
sequence stars having the potential to form planetary systems is now well 
established \citep{Koerner01}. However,  unknown is an accurate estimate of 
the fraction of single stars with disks that have produced planetary 
systems. If not all such systems have planets, why then do only some 
accretion disks form planets and others do not? Theoretical models of 
particle aggregation show that if particles can grow from submicron to mm 
to cm in size, then the formation of planetesimals is possible in the time 
before the disk dissipates \citep{Habi99}. The problem 
remains to understand how 
grains condense from nebular gasses, and how relic interstellar grains 
survive and are modified by their transport in the disk.  These pristine 
submicron size grains, be they nebular condensates or relic interstellar 
grains, then grow into mm size particles. However, investigating how 
grains grow is complicated by the fact that most of the grain processing 
probably occurs in the hot mid-planes of these disks \citep{Bell2000}, 
hidden from view by the optically thick disk photospheres. If grains are 
lofted above the photosphere by disk processes such as winds, turbulent 
convection, or changes in vertical structure, the evolution of dust can be 
investigated by observing the properties of the small ($\leq 1$~\micron) 
grains in the disk surface layer or atmosphere. 

In particular, observing and modeling the 10~\micron \ silicate resonances 
in nearby (d~$\leq 500$~pc), young (age~$\leq 5$~Myr) Herbig Ae/Be (HAEBE) 
stars, the sites of possible on-going planetary formation \citep{WW98}, 
establishes the silicate mineralogy, Mg content of the grains, the degree 
of crystallinity vs.\ amorphousness, grain porosity, and grain size 
distributions.  These grain properties can be linked to conditions in 
protoplanetary disks and constrain theoretical models of protoplanetary 
disk evolution. For example, amorphous silicates commonly detected in 
solar system comets are considered to be interstellar relic grains 
\citep{LiGre97} and probably are represented by Glasses Embedded with 
Metals and Sulfides (GEMS) in interplanetary dust particles, IDPs 
\citep{Brownl2000}.  Conversely, Mg-rich crystalline silicates are thought 
to be either pristine solar nebula condensates that condensed at $\sim 
1450$~K or amorphous silicates annealed into crystals at temperatures 
$\gtrsim 1000$~K in the inner hot zones of the accretion disk.  If these 
silicate crystals are formed in the hot inner zones of the disk, then the 
crystals must migrate to large radial distances in order to be 
incorporated into comets and outer nebula protoplanetesimals.  However, 
the hot temperatures required to form the silicate crystals could have 
occurred in the early phases ($\sim 300,000$~yrs) of the solar nebula in 
the disk \citep{BM02} or in nebula shocks in the $5 - 10$~AU region, 
obviating the need for radial transport of the crystals to comet forming 
zones.  Indeed, Mg-rich silicate crystals within IDPs are detected through 
cometary IR spectra \citep{Wooden2000a,Wooden2000b}. Crystalline silicates 
also have been spectroscopically identified in pre-main sequence HAEBE 
stars including HD~100546 \citep{Malf98} and in $\beta$~Pic 
\citep{Pant97}, a debris 
disk system continually repopulated with dust grains via comet-comet 
collisions and cometary accretion events. The HAEBE stars (the high mass 
[$2 - 10$~M$_{\odot}$] analogs of T Tauri stars [$\sim 1$~M$_{\odot}$]) 
are the probable progenitors of the $\beta$-Pic stars since both have a 
common occurrence of `exocometary activity,' are similar in spectral type, 
and have an apparent overlap in Hipparcos ages \citep{Crifo97}. 

In this work, we examine the thermal emission from three Herbig Ae/Be 
stars of similar spectral type: HD~150193 (A1V), HD~100546 (B9V), and 
HD~179218 (B9).  The stellar ages of the objects are: $2 \times 10^6$~yrs, 
$10 \times 10^6$~yrs, and $0.5 \times 10^6$~yrs for HD~150193, HD~100546 
and HD~179218, respectively \citep{vda98}. We chose these objects for study 
because they are isolated (not belonging to any know star forming region), 
have little to no active accretion, possess possible processing 
circumstellar disks, and have varying degrees of silicate crystallinity 
\citep{Meeus01}. We assemble and model SEDs for each system using a passive 
reprocessing  circumstellar disk to constrain the amount of crystalline
silicates in the inner and outer parts of the disk, and flaring of the disk 
in each of the systems. 

\section{OBSERVATIONS}
\label{observations}

Spectrophotometry of HD~150193, HD~100546, and HD~179218 was obtained 
using the NASA Ames Research Center {\it HI\/}gh efficiency {\it F\/}aint 
{\it O\/}bject {\it G\/}rating {\it S\/}pectrometer (HIFOGS:  
\citet{Witteborn91}).  The spectrum of HD~179218 was obtained on the 2.34-m 
telescope at the Wyoming Infrared Observatory (WIRO) using 32-bit FORTH 
telescope software.  The spectra of HD~150193 and HD~100546 were obtained 
on the 4-m Blanco telescope at the National Optical Astronomical 
Observatories Cerro Tololo Inter-American Observatory (CTIO) in Chile.  
Table~\ref{tab:obshf} summarizes the HIFOGS observations used in this 
work. 

\begin{figure}
\epsscale{0.5}
\plotone{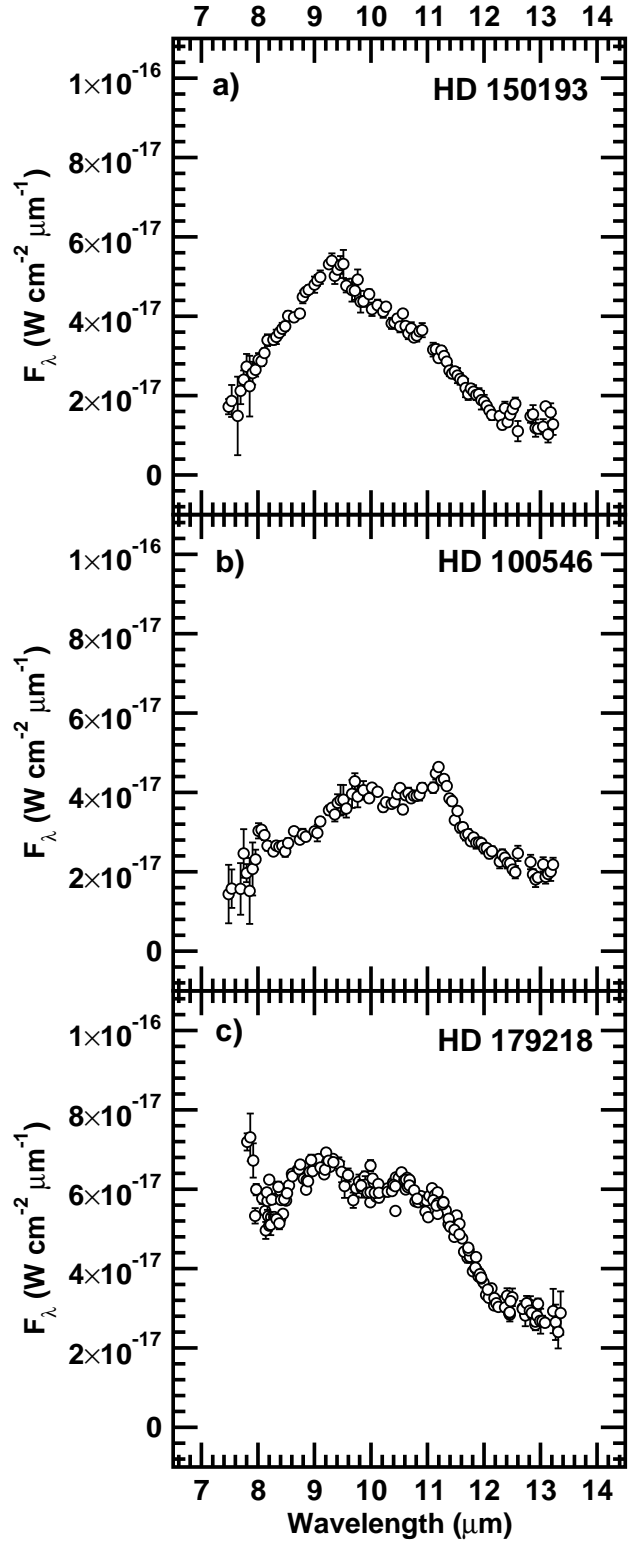}
\caption{HIFOGS spectra of HD~150193, HD~100546, and HD~179218.
\label{fig:hifogs} }
\end{figure}

The HIFOGS is a grating spectrometer with a resolution of $\Delta \lambda 
\simeq 0.045$~\micron\ per detector, yielding a resolving power of $R = 
\lambda /\Delta\lambda \simeq 350 - 180$ covering the $7.5 - 
13.4$~\micron\ spectral region.  A $3^{\prime\prime}$ diameter circular 
entrance aperture was used for each observation at each observing 
facility.  This aperture corresponds to a radial size of 448~AU, 309~AU, 
and 732~AU for HD~150193, HD~100546 and HD~179218, respectively. Standard 
IR observing techniques of chopping and nodding were employed, with a chop 
frequency of 7~Hz and with a chop throw of $30^{\prime\prime}$ in the E-W 
direction.  Absolute calibrated flux spectra of the standard stars used in 
the data reduction were taken from the UKIRT CGS 3 and HIFOGS measurements 
compiled by \citet{Coh96}. Ratios of two or more measurements for each 
standard at different air masses were fitted to ratios of atmospheric 
transmissions (using ATRAN software, \citet{Lord93}), for a precipitable 
water vapor of 4.5~mm for both WIRO and CTIO to verify the wavelength 
calibration and the atmospheric transmission corrections as functions of 
airmass.  Each HIFOGS spectrum was flux calibrated by dividing the 
measured spectrum by the standard star, multiplying by the corresponding 
star's flux spectrum \citep{Coh96}, and multiplying by the inverse ratio of 
computed atmospheric transmission spectra.  All observations presented 
here were obtained under photometric conditions. The spectra are shown in 
Fig.~\ref{fig:hifogs}. 

Complete Nyquist sampling of the 10~\micron\ spectral signatures requires 
two HIFOGS grating settings.  We were able to achieve Nyquist sampling for 
the object HD~179218.  The other two objects, HD~150193 and HD~100546, 
were observed at a single grating setting.  Observations of HD~179218 on 
1998 June 23~UT, June 27~UT, and June 30~UT were obtained at the first 
grating setting that spanned $7.811 - 13.356$~\micron.  The 1998 June 
24~UT and June 25~UT observations were obtained at a second grating 
setting that was shifted such that the wavelength interval of the 
detectors was half way between the wavelength interval of the detectors in 
the first grating setting.  The second grating setting spanned a 
wavelength range of $7.519 - 13.122$~\micron.  The spectra from each 
observational epoch are shown in Fig.~\ref{fig:hifogs}.  Spectra for each 
grating setting were scaled to the spectrum obtained during the best 
photometric conditions (1998 June 23~UT and 1998 June 25~UT for the first 
and second grating settings, respectively) and combined by computing a 
statistically weighted average.  The scaling factors for each spectrum are 
listed in Table~\ref{tab:obshf}.  The resultant HIFOGS spectra for 
HD~179218 at each grating setting were subsequently interlaced to produce 
the final spectrum (spanning $7.528 - 13.356$~\micron).  Therefore, for 
HD~179218, there are approximately twice as many data points between 
7.5~\micron\ and 13.4~\micron\ than for HD~100546 and HD~150193. No 
scaling factor was used to interlace the spectra from the two grating 
settings. 

We also obtained 1.2 -- 23.0 \micron\ IR broad- and narrow-band 
photometric observations of HD~179218 on 2001 April 30.42~UT using the 
University of Minnesota (UM) Mount Lemmon Observing Facility (MLOF) 1.52~m 
telescope and the UM multi-filter bolometer \citep{Hanner90}. Observations 
where conducted using 9.33\arcsec\ circular aperture, a 29\arcsec\ N-S 
nod, and a 7.5~Hz chop-frequency. In addition to standard near-IR 
broad-band filters, the bolometer filter set also included the six 
Infrared Telescope Facility (IRTF) narrow-band 7--13~\micron\ ``silicate'' 
filters \citep{Tokunaga86}.  Multiple observations of the photometric 
standard star $\alpha$~Boo (= HD~124897) at a variety of air masses during 
the night established the photometric zero point over the entire spectral 
range of the bolometer. The photometric systems, magnitude scales, and 
absolute flux calibrations for the bolometer are given by \citet{Mason01} 
and \citet{rdg97a,rdg97b}. Standard extinction values of 0.2 
magnitudes per air mass (mag AM$^{-1}$) were used for filters $[J] - [L]$ 
and 7.9--12.6 \micron, 0.3~mag~AM$^{-1}$ for filter $[M]$, and
 0.5~mag~am$^{-1}$ at 18.3~\micron. The use of standard extinction values 
introduces no more than a few percent uncertainty in the derived 
photometry, as observations of HD 179218 where obtained at $\leq 1.2$ air 
masses. Observed magnitudes corrected for atmospheric extinction, are 
summarized in Tables~\ref{tab:obsbolo}. 

\subsection{Assembly of Data Sets for SEDs}
\label{sec:assembly}

Data from the European Space Agency Infrared Space Observatory ({\it ISO}) 
Short-Wavelength Spectrometer (SWS) and the Infrared Astronomical Observatory
(IRAS) are used to expand the wavelength coverage of the spectral energy 
distribution (SED) of each of our three HAEBEs.  
The {\it ISO} SWS observations were conducted using the 
SWS01 scanning mode (Astronomical Observing Template, referred to as AOTs) 
covering the entire 2.4 -- 45~\micron\  SWS wavelength range. 
Each object has been observed at a different scanning speed; in particular,
data have been taken at ``speed 4'' (HD100546), ``speed 3'' (HD179218),
and ``speed 1'' (HD150193), corresponding to raw resolution of
approximately $\lambda/\Delta\lambda \simeq $ 1400, 800, and 400,
respectively.

Since the HAEBEs were faint and the standard pipeline process did not 
adequately apply a dark subtraction to the data points, we have 
reprocessed the data from a raw level to a fully processed stage using the 
Observer's Spectral Interactive Analysis Package (OSIA v2.0) and the {\it ISO} 
Spectral Analysis Package (ISAP v2.1). Each of the 12 bands into which the 
SWS01 is divided have been processed separately; dark current subtraction, 
scan direction matching and flat-fielding have been applied interactively 
to each band. Using ISAP, we sigma clipped outliers and averaged the data 
of all 12 detectors for each AOT band, retaining the instrument 
resolution.  For those bands affected by fringes, we applied the defringe 
routines incorporated in ISAP. 

For extended sources a flux discontinuity may occur between band limits as 
a consequence of a change in aperture size at some band edges. Our three 
objects show very little flux jumps at the detector band limits, and in 
most cases, we did not have to normalize the segments to their neighbors 
when we combined the 12 spectral segments into a single spectrum. For 
HD~179218 we use an additive term to normalize the last two segments 
($19.5 - 45$~\micron) to the flux in the previous band. Since we have {\it ISO}
SWS data from objects that have been taken at different scanning speeds and, 
consequently, have different resolutions, we have produced the final 
spectrum for each object rebinning the data to a bin size that corresponds 
to a spectral resolution of $\lambda/\Delta\lambda \simeq 400$. 

The {\it ISO} SWS spectra are scaled to the HIFOGS spectrophotometry using 
the following procedure: 1)~the resolution of the {\it ISO} SWS spectrum is 
degraded to match the wavelength scale of the HIFOGS spectrum; 2)~a scaling 
factor is calculated for each data point (approximately 30 points for 
HD~100546 and HD~150193, and 60 points for HD~179218) in the high S/N region
between 8 and 10~\micron; 3)~a statistically weighted average of the 
scaling factors is computed; and 4)~the averaged scaling factor is applied to 
the entire original, undegraded {\it ISO} SWS spectrum.  The scaling factor 
of the {\it ISO} SWS spectra to the HIFOGS spectrophotometry for each object 
is listed in Table~\ref{tab:sedscale}.  

In addition to spectra from {\it ISO} SWS, we use IRAS flux values 
from \citet{Oudm92} for each of the three objects.  We apply
the method outlined in the IRAS Explanatory Supplement (1988) to
produce color corrected, monochromatic flux points.  We assume the IRAS fluxes 
are larger due to the larger IRAS aperture.  First, the 12~\micron\ IRAS 
data point is scaled to the HIFOGS spectrum.  Then, the resulting 
scaling factor is applied to each of the other three IRAS data points 
(Table~\ref{tab:sedscale}).

\begin{figure}
\epsscale{0.8}
\plotone{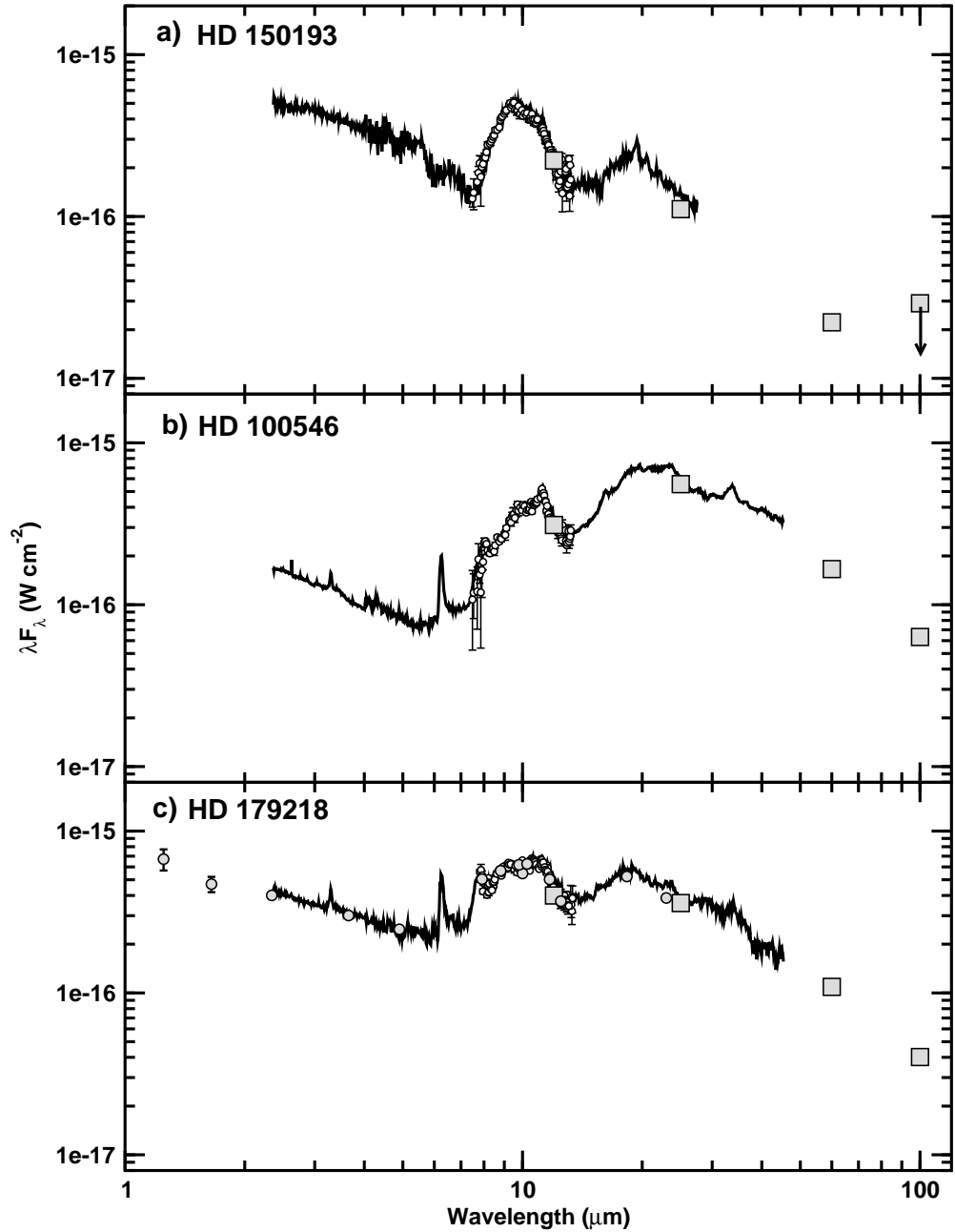}
\caption{Assembled SEDs for $a)$~HD~150193, $b)$~HD~100546, and 
$c)$~HD~179218.  Scaled to the HIFOGS spectrum 
({\it open circles}) for each SED is the {\it ISO} SWS spectrum 
({\it black line}), and the IRAS photometric points 
({\it gray squares}) corresponding to each object.  The 100~\micron\
point for HD~150193 (panel $[a]$) is an upper limit and not used in the model
fitting.
\label{fig:sed} }
\end{figure}

Finally, we added the IR photometric points from MLOF to the SED of 
HD~179218. The near-IR points were corrected for interstellar reddening by 
assuming the extinction law of \citet{Mathis00} and the published 
extinction for HD~179218 of $A_v = 1.6$.  The scaling factor applied to 
the MLOF data to match the HIFOGS spectrum of HD~179218 is listed in 
Table~\ref{tab:sedscale}. The final assembled SEDs for all three objects 
are shown in Fig.~\ref{fig:sed}. 

The factors needed to scale the {\it ISO} SWS data to the HIFOGS data 
arise due to a difference in aperture size and distance to the object.  
The effective aperture size of {\it ISO} SWS is 9.44\arcsec, and for 
HIFOGS is 3\arcsec. Based on millimeter studies, the diameter of the disk 
around HD~179218 is $230 - 450$~AU \citep{ms2000} and for HD~150193 is 
$\sim 250$~AU \citep{ms1997}.  The distance to HD~179218 (based on Hipparcos 
data) is 244~pc and for HD~150193 is 149~pc.  This results in projected 
diameters of the {\it ISO} SWS and HIFOGS apertures on HD~179218 of 
2302~AU and 732~AU.  Similarly, the projected diameters on HD~150193 are 
1409~AU and 448~AU.  Therefore, the apertures contain the millimeter 
measured disk sizes resulting in {\it ISO} SWS to HIFOGS scaling factors 
close to 1 for these two objects.  Conversely, the distance of HD~100546 
is 103~pc yielding a projected aperture diameter from {\it ISO} SWS and 
HIFOGS as 973~AU and 309~AU, respectively.  Although there are no 
millimeter measurements of HD100546, HST Chronographic Imaging reveals a disk
approximately 760 -- 1030~AU in diameter \citep{Auge2001}, \citep{Grad2001}.
Therefore, significant mid-IR flux may be contained in the {\it ISO} SWS 
aperture which would account for the comparatively large scaling factor needed 
to scale {\it ISO} SWS to HIFOGS for HD~100546. 

\section{Passive Disk Model}

\begin{figure}
\epsscale{0.8}
\plotone{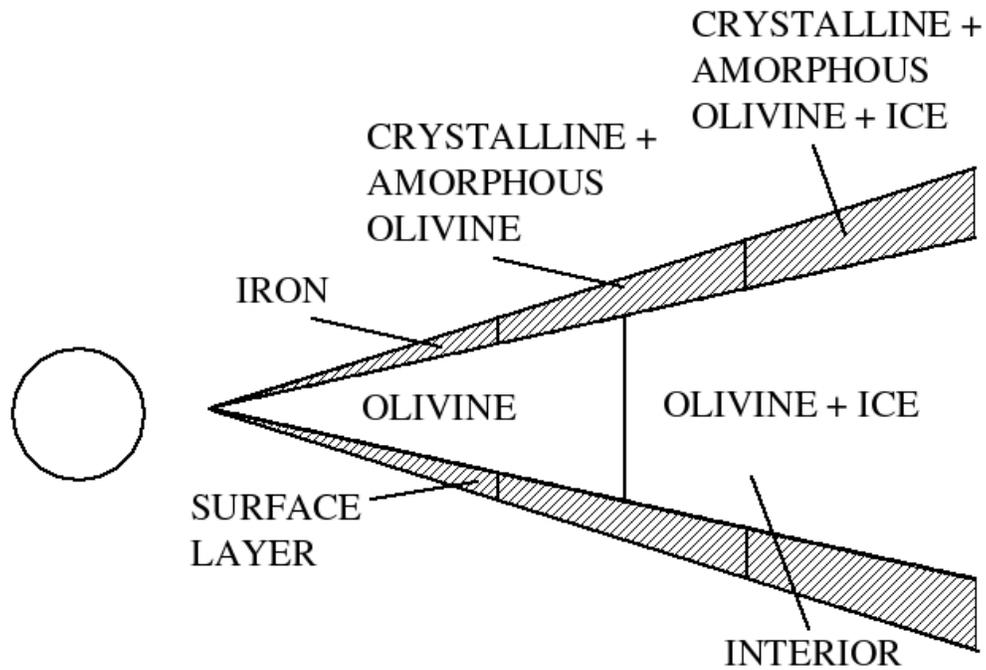}
\caption{Cross-sectional schematic of our disk model, adapted from Fig.\ 1 
in CO1.  Identified in the sketch is the relative location of each mineral
species.  The {\it vertical lines} indicate the radial zones in the disk in 
which a particular mineral dominates the emission, and the {\it dashed lines}
indicate the separation between the disk interior and the disk surface
layer.  The disk interior and disk surface layer ({\it hash lines}) are
indicated.
\label{fig:sketch} }
\end{figure}

We use a modified version of the radiative, hydrostatic models of 
passively irradiated circumstellar disks developed \citet{chigol97} (hereafter 
CG97) and updated by \citet{chiang01} (hereafter C01).  The CG97 disk is 
composed of two parts: an optically thin surface layer and an optically 
thick interior (Fig.~\ref{fig:sketch}).  The surface layer is directly heated 
by light from the 
central star.  Half of the reprocessed emission from the surface layer 
escapes into space while the other half radiates onto the disk interior.  
The emission from the disk surface layer heats the cooler disk interior.  
Since the disk is in hydrostatic equilibrium, the disk flares with 
increasing distance from the central star.  The amount of disk flaring is 
influenced by the temperature of the interior.  Flaring of the disk 
results in more starlight hitting the disk at larger distances from the 
central star as compared to an otherwise geometrically flat disk.  C01 
updated the CG97 model by adding a range of grain sizes and incorporating 
a simple mineralogy  through the use of laboratory determined optical 
constants, including water ice, amorphous olivine, and metallic Fe.  
Although the two-layered disk approach is simple, when compared to more 
detailed vertical structure models, the CG97 model was found to be a 
robust model for computing SEDs from HAEBEs \citep{Dull02}. 

In this work, we use the updated model of C01 with our own enhanced 
feature of adding Mg-pure crystalline olivine grains (forsterite) into the 
optically thin surface layer. We also are able to vary the 
crystalline-to-amorphous silicate ratio in the inner and outer radial regions 
of the disk.  The details of the standard disk model are explained in CG97 and 
C01; however, we will highlight the disk and dust parameters here.  

Disk parameters include the disk surface density at 1~AU 
($\Sigma_{\circ}$), the outer disk radius ($r_{\circ}$), and the height of 
the disk photosphere in units of the gas scale height ($H/h$). Dust grain 
parameters include grain mineralogy, slope of the grain size distribution 
($q$), maximum grain size ($a_{max}$), and sublimation temperature 
($T_{sub}$). The shape of the SED in the region covered by our data sets 
(2 -- 45~\micron) are largely unaffected by changes in the input 
parameters.  The dust parameter which most affects the shape of the resonance 
features is the slope of the grain size distribution, $q$ (C01).
Changes in disk radius, disk surface density, and maximum 
grain size for the interior and the surface layer mostly affect the shape 
of the SED at wavelengths longer than 50~\micron\ (C01; Creech-Eakman et 
al.\ 2002).  The one disk parameter that produces the most change at 
wavelengths less than 50~\micron\ is $H/h$, which can be used as a measure
of dust settling in the disk (C01).  However, as noted in C01, 
$H/h$ is not a fixed parameter and should be calculated self-consistently at
each radius in the disk.  We follow the procedure of C01 by keeping $H/h$
constant throughout the disk and using it as an indicator of dust settling 
in the disk.  Another unknown parameter expressed in terms
of $H/h$ may actually represent the settling of dust in the disk.
Finally, the relative mass fraction of the minerals in the disk and surface 
layer will influence the SED in the mid-IR. 

To model the emission from dust particles, we use the method of C01 and 
assume that the disk is composed of a size distribution of isolated 
spherical grains of discrete mineralogy; i.e., we ignore gas as a source 
of opacity.  The specific mineralogy is addressed in the next section 
(\S\ref{sec:mins}).  The opacity of the grains is computed by using Mie 
Theory.  For grains coated with a water ice mantle, Mie-G\"uttler Theory 
is used to compute the opacity.  The grain size distribution in the disk 
surface layer and in the disk interior are equal, and unchanged in each of 
our models.  The central stars are modeled as simple blackbodies.  The 
radius, temperature, and luminosity of the central stars of our three 
objects are taken from the compilation by \citet{vda98}.  We use the 
luminosity to calculate the radius of each of the stars using the 
relation: $L_{\star} (\rm{ergs~cm}^{-2}~\rm{s}^{-1}) = 4\pi \sigma R_{\star}^2 
T_{\star}^4$. 

Two final assumptions are made when using the C01 model.  The first is 
that we are viewing the disk face-on.  The stars discussed here have low A$_v$ 
($A_v = 0.28$ for HD~100546; $A_v = 1.27$ for HD~179218; and $A_v = 1.6$ for
HD~150193), therefore, the disk is not inclined at such an angle as to 
significantly block the starlight.  CG99 showed that until the disk is 
inclined at such an angle as to block the starlight ($i \sim 45^{\circ}$), 
the SED is mostly unaffected by disk inclination.  Therefore, 
we use the approximation that the disks are face on for all three of our 
objects.  The second assumption is that any accretion in these systems
is negligible and does not significantly contribute to the measured
emission from these systems.  By making such an assumption, we are placing an
upper limit on the contribution of flux from the disk alone on the measured
emission from each of these systems.

Flux from the central star, disk interior and disk surface are
co-added to produce the model SED.  The relative amount of amorphous
to crystalline silicates in the disk surface layer is adjusted to produce the
best fit within a $2\sigma$ (95\% confidence) level \citep{Press92}.

\subsection{Dust Mineralogy}
\label{sec:mins}

The mineralogy used by C01 is a good representation of the basic 
minerals extant in young disk systems \citep{Malf98,Bouwm2000}.  We 
maintain the mineralogy of C01, using amorphous olivine, metallic Fe, and 
water ice.  However, we have expanded the mineralogy by including Mg-pure
crystalline olivine grains in the disk surface layer.  Crystalline olivine 
has been discovered to be a very important component of the dust in 
circumstellar disks and solar system comets.  In an attempt to quantify 
the amount of crystalline olivine grains, we make this important step of 
including them in the C01 passive disk model.  Table~\ref{tab:mins} lists 
the minerals used in the modeling and the reference for the indices of 
refraction. 

It is difficult to model the thermal emission from crystalline silicate 
grains \citep{yfh99}. Many authors choose to use a form of a continuous 
distribution of ellipsoids while others use direct comparisons from 
laboratory transmission experiments.  In this work, we choose to use the 
methods outlined by \citet{Fabi01} who calculate the emission from 
ellipsoids by elongating the crystals along one of the three crystalline 
axes.  Based on the location of the resonance peaks in the HIFOGS and {\it 
ISO} SWS spectra of HD~150193 and HD~179218, we use an axis ratio of 
10:1:1 to compute the optical efficiencies (Q$_s$) of the crystals.  The 
crystalline olivines are not coated with ice since to do so requires 
mixing theory which eliminates the distinctive crystalline resonances.  We 
compute the thermal emission from the crystals in the disk surface layer 
of each object. We make the simplified assumption that the crystals are at 
all radii in the disk surface layer. This assumption may not be valid 
since ice should coat the crystals out at larger disk radii.  However, the 
measured crystalline resonances seem to originate from bare crystalline 
grains.  Finally, we vary the ratio of amorphous to crystalline silicates 
to find the best model fit to the SEDs. 

We note that in addition to solid-state emission features from
silicates, HD~100546 and HD~179218 exhibit emission from polycyclic
aromatic hydrocarbons (PAHs) \citep{Meeus01}.  We do not attempt to model 
the emission from these molecules here, but will examine them in a future 
paper.

\section{RESULTS}
\label{sec:results}

\begin{figure}
\epsscale{0.75}
\plotone{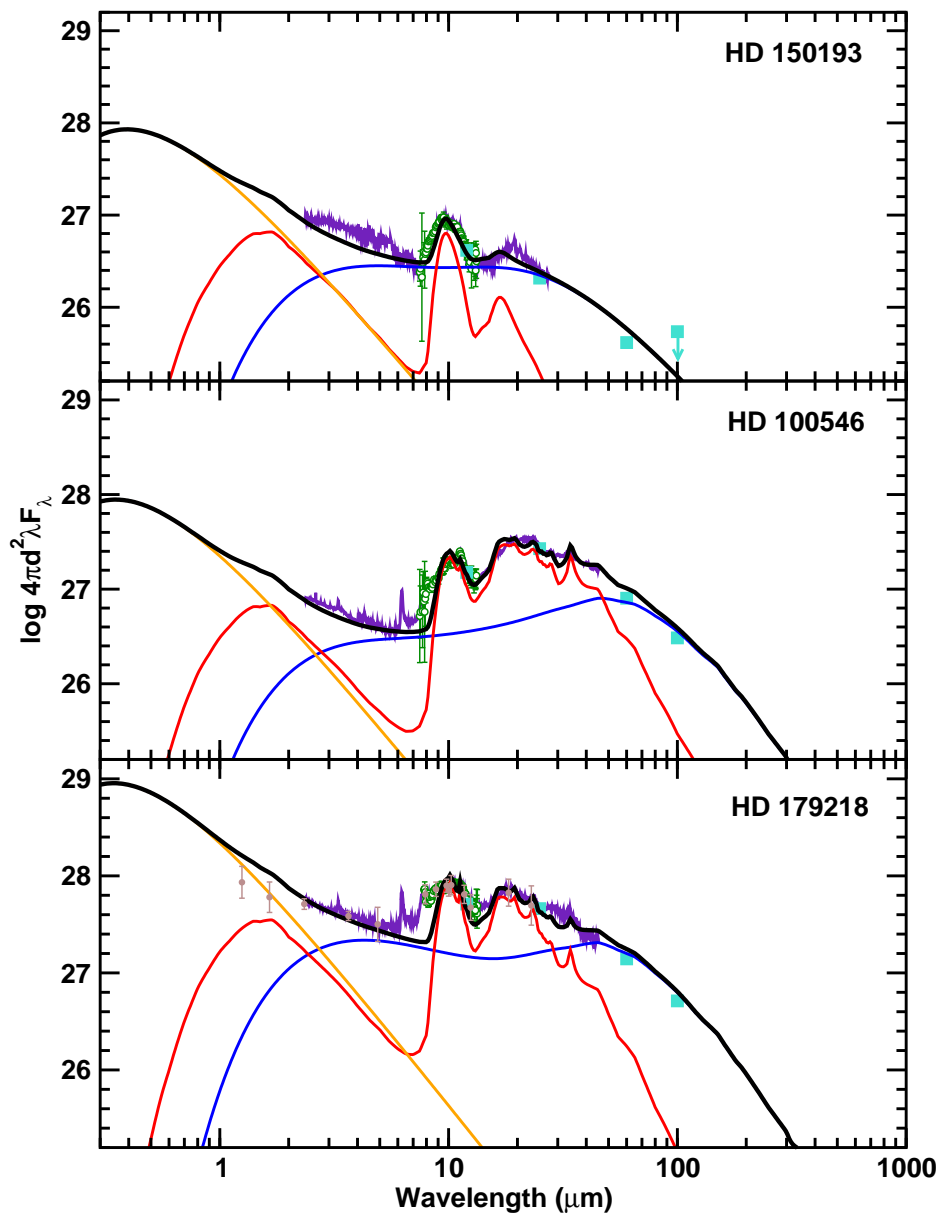}
\caption{C01 model computed for three Herbig Ae stars: 
HD 150193, HD 100546 and HD179218.  The disk interior (blue line), disk 
surface (red line), and stellar blackbody (orange line) are co-added to 
produce the model SED (black line).  The model SED is compared to the 
assembled data sets including: ISO SWS spectra (indigo line), HIFOGS spectra 
(green circles) and IRAS photometry points (turquoise squares).  HD 179218 
also has MLOF photometry points (brown circles). 
\label{fig:model} }
\end{figure}

\begin{figure}
\epsscale{0.50}
\plotone{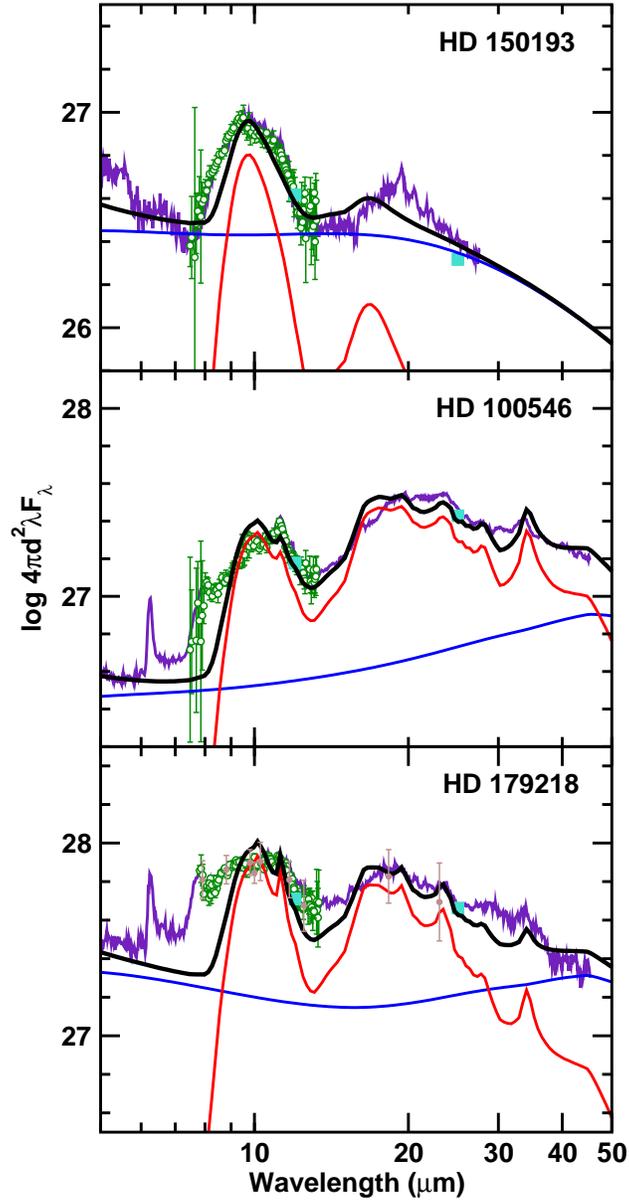}
\caption{Zoom in on the C01 model fits of the crystalline olivine features for 
the three Herbig stars: HD 150193, HD 100546 and HD179218.  The colors are the 
same as in Fig.~\ref{fig:model}.
\label{fig:10model} }
\end{figure}

Our modeling results are displayed in Figs.~\ref{fig:model} and 
\ref{fig:10model} and summarized in Table~\ref{tab:fitparams}.  To produce the 
best-fit model to the SEDs, HD~100546 and HD~179218 are both modeled with a 
disk 150~AU in radius, while HD~150193 is modeled with a disk 5~AU in radius.  
The best fit SEDs for 
HD~100546 uses a scale height of $H/h = 4$, a scale height of $H/h = 3$ for 
HD~150193, and a scale height of $H/h = 1$ for HD~179218.  This means that 
at a radius of 5~AU the disk around HD~100546 flares about 80\% more than the 
disk around HD~179218, and about 25\% more than the disk around HD~150193.  
The current model fits suggests that the measured emission from 
HD~150193 is dominated by warm dust in a flaring disk close to the 
central star, and that the emission from grains in an extended disk ($> 5$~AU) 
are relatively small compared to the other two objects.  We should note that 
for HD~150193, a larger disk of radius 50~AU with a scale height of $H/h = 1$ 
can be used to match the contrast of the measured silicate resonance features 
in HD~150193, but the calculated flux at longer wavelengths exceeds the 
measured IRAS data point at 60~\micron\ by a factor of 4.  Therefore, we are 
able to match the long wavelength photometry and the contrast of the silicate 
resonance features by reducing the radius of the disk and increasing the scale 
height.

The qualitative crystalline content of these three HAEBEs has previously been 
reported by \citet{Meeus01} with HD~100546 and HD~179218 showing evidence of 
crystalline silicates, while the emission from HD~150193 is dominated by 
amorphous silicates.  For the two 
objects with crystalline silicates, we find that a better fit is produced to 
the observed SEDs of the objects if we use a model in which a higher ratio of 
crystalline-to-amorphous silicates is located in the inner regions of the disk
($\leq 5$~AU) compared to the outer regions of the disk (5 -- 150 AU).  The 
inner region of HD~100546 has 37\% higher crystalline-to-amorphous silicate
ratio compared to the outer region.  This is contrary to the findings of 
\citep{Bouwm2003} (hereafter, BdKDW) who used a spherical shell model to 
calculate a factor of almost 10 higher fraction of crystalline silicates in 
regions greater than 10 AU.  The inner region of HD~179218 has 84\% higher 
crystalline-to-amorphous silicate ratio compared to the outer region.  
Our model results are consistent with the recent findings of \citet{vanb2004}.
\citet{vanb2004} observed three Herbig Ae stars (HD~142527, HD~163296, and
HD~144432) using the Mid-Infrared Interferometric Instrument on the Very
Large Telescope Interferometer.  In all three of their objects, they found 
evidence for a larger fraction of crystalline silicates in closer to the star 
(1 -- 2~AU) compared to the outer region of the disk (2 -- 20~AU).

Finally, all three of our objects are best fit using a grain size distribution 
with a slope of $q = 3.5$ for both the optically thick inner disk and for the 
optically thin surface layer.

\section{DISCUSSION}
\label{sec:discussion}

It is difficult to make any statistically significant conclusions about disk 
evolution from modeling only three objects.  However, we can make some 
interesting observations based on our results.  \citet{Meeus01} defined 
two groups of Herbig Ae/Be stars based on the shape of their SEDs.  Group I 
objects exhibit near-IR and far-infrared (far-IR) emission, with and without 
silicate emission features (denoted Groups Ia and Ib, respectively).  Group II 
objects exhibit near-IR emission and silicate emission features, but much less 
far-IR emission compared to the objects in Group~I.  HD~179218 and HD~100546 
are considered Group Ia objects and HD~150193 is considered a Group II object 
under this scenario \citep{Meeus01}.  

\citet{Meeus01} suggest a disk geometry interpretation of their groupings.
Group I objects have a geometrically thin, optically thin 
inner region with a flaring outer region.  Group II objects contain an 
optically thick inner region which puffs up, shielding the optically thin, low 
mass, outer disk 
region from stellar flux, thereby preventing it from flaring.  
Qualitatively our modeling results match these groupings.  HD~150193 is 
modeled with a relatively small (5~AU), low mass disk, which suppresses 
the amount of emission from ice coated silicate grains contributing to the 
emission around 45~\micron.  HD~179218 and HD~100546 are modeled with 
relatively large disks. HD~100546 is modeled with a disk that flares more 
than that of HD~150193 and HD~179218.  We find it interesting that the 
disk of HD~179218 flares less than the disk of HD~100546, even though the 
two are both Group I sources.  At least on the quantitative level, the two 
disks differ. 

Although our model fits show no evidence for grain growth (all three SEDs are 
modeled with the same grain size distribution) there is a difference in the 
radial distribution of crystalline silicates.  A possible evolutionary 
scenario is one in which when circumstellar disks form around young stars, 
they are thought to be primarly composed of amorphous ISM grains 
\citep{Wood2004}.  As the disk evolves, crystalline silicates are condensed 
\citep{Gros1972} or annealed \citep{Riet2002} in the inner radial regions of 
the disk, either through heating \citep{BM02} or through shocks in the
disk \citep{Hark2002}.  The crystals are then transported to the outer 
regions of the disk \citep{Cuzzi93}.  Therefore, from this scenario, we 
can conclude that the disk around HD150193 is the least evolved since it does 
not contain any crystals.  This is followed by HD179218 which has a large 
crystalline-to-amorphous in the inner radial region compared to the outer 
radial region.  Finally, this is followed by HD100546 which has a slightly
lower crystalline-to-amorphous silicate ratio at inner disk radii and a
larger ratio at large disk radii compared to 
HD179218.  Such a scenario is supported by the findings of \citet{vanB2003}
who find that the shape and strength (band over continuum) of the 10~\micron\ 
silicate feature are correlated.  Strong features (from small grains)
show mostly amorphous type silicate grains, while weaker, flatter features
show evidence of crystalline silicates (grain processing).

It should be noted that there may be evidence of grain growth in both HD~179218
and HD~100546 as evidenced by the flatness of their 10~\micron\ silicate 
features.  We have not been able to adequately model ``by eye'' the shape of
the feature with the current model.  Attempts to use a more shallow grain
size distribution and/or truncate the lower end of the grain sizes did not
result in better fits (either by eye, or statistically) to the 10~\micron\
feature.  A more diverse mineralogy combined with a radially varying grain
size distribution may improve the fits to the 10~\micron\ spectral feature.
However, such models are beyond the scope of this paper.

\subsection{Comparison with Other Modeling Efforts}

Other authors have modeled the three objects presented here.
As stated earlier, BdKDW modeled the SED of HD~100546 using
a optically thin spherically symmetric dust distribution (i.e., no optically 
thick disk component).  BdKDW find that
the the amount of crystals used to fit the SED increased with radial
distance from the star.  This contrasts with our model which shows that
there needs to be a higher crystalline-to-amorphous silicate ratio in
closer to the star, in agreement with radial mixing models.  BdKDW
also finds that HD~100546 has a large, flaring disk, consistent with our
model results.

\citet{Domi2003}, using another variation of the CG97 model \citep{Dull01} 
(hereafter DDN), used a simple mineralogy of grains 0.1~\micron\ in size.  
They modeled HD~150193 with a disk smaller than that of the other two and  
found that the emission feature in HD~100546 is best fit with larger sized 
grains than the other two objects.  While we can not conclusively make a 
quantitative statement about the grain size in HD~100546, our results are 
qualitatively similar with \citet{Domi2003} for HD~150193.  However, there are
some important differences between our model results and those of
\citet{Domi2003}. They find that their fits for these three objects are 
best when the surface density increases with distance from the star 
(instead of decreasing as we have assumed here).  They interpret these 
results as evidence for a ``gap'' in their disks, possibly produced by a 
planet or other large body.  We do not find such evidence for an 
increasing surface density in our model calculations.  Furthermore, based 
on DDN, it is extremely difficult to disentangle the degeneracy of the DDN 
model when simultaneously varying the surface density, the slope of the 
surface density with stellar distance, and the radial size of the disk.  
Therefore, while both of our results are intriguing, it underlines the 
need (as pointed out in Dominik) for high spatial resolution imaging of 
these objects in the mid-IR to resolve these modeling differences. 

\section{CONCLUSIONS}
\label{sec:conclusions}

In this paper we constructed mid-IR SEDs of three Herbig Ae/Be stars using 
data from the HIFOGS, {\it ISO} SWS, IRAS, and MLOF photometry.  We modeled 
the SEDs using the simple two-layer radiative transfer disk model of CG97 
and C01 with the added feature of calculating the emission from Mg-pure
crystalline olivine grains.  Our findings are highlighted as follows: 

1)~HD~150193 has thermal emission arising from a relatively small 
($\simeq 5$~AU in radius) flared ($H/h = 3$) disk.

2) HD~179218 has thermal emission arising from a large ($\simeq 150$~AU
in radius) less flared ($H/h = 1$) disk.

3) HD~100546 has thermal emission arising from a large ($\simeq 150$~AU
in radius) more flared ($H/h = 4$) disk.  At 5~AU, the disk around HD~100546
flares 80\% more than the disk around HD~179218 and 25\% more than HD~150193.

4) All three objects were modeled using a power law grain size distribution 
with a slope of $a^{-3.5}$ for both the optically thick inner region and
the optically thin surface layer.

5) HD~150193 shows no evidence of emission from crystalline silicates
while HD~179218 and HD~100546 both show emission from Mg-rich crystalline
olivine grains.

6) The SED of HD~179218 was best fit using a crystalline-to-amorphous
ratio 84\% larger in the inner radial regions of the disk ($\leq 5$~AU)
compared to the outer regions.

7) The SED of HD~100546 was best fit using a crystalline-to-amorphous
ratio 37\% larger in the inner radial regions of the disk ($\leq 5$~AU)
compared to the outer regions.

In circumstellar disks, grain growth and crystallization may require both 
sufficient disk mass (i.e., size) and scale height.  Scale height may be 
a signature of turbulent movement in the disk which could aid grain growth 
and transport \citep{Cuzzi93}. More disks need to be fitted with similar 
model structures to increase the sample to make more sense of the links 
between grain properties and disk structure. 

\acknowledgements
DEH and CEW acknowledge partial support for this work from from NSF Grant 
AST-0205814. In addition, DEH and DW acknowledge support from the NASA Ames 
Research Center Director's Discretionary Fund. The authors thank 
the day-crew and staff of NOAO CTIO for their assistance at the Blanco 4-m 
(in particular Ron Probst) and James E. Lyke for his assistance with the 
UM MLOF observations.  Finally, the authors wish to thank an anonymous 
referee for providing useful comments on the paper.

\newpage

\clearpage

\begin{deluxetable}{lccccccc}
\tablewidth{0pc}
\tablecaption{HIFOGS OBSERVATIONAL SUMMARY \label{tab:obshf}}
\tablehead{
 & \colhead{UT} & \colhead{UT Time} & \colhead{Integ.} & 
\colhead{Airmass} & \colhead{Flux} & \colhead{Airmass of} & \colhead{Scaling}\\
\colhead{Object} & \colhead{Date} & \colhead{(hr:min)} & \colhead{(min)} & 
\colhead{Range} & \colhead{Standard} & \colhead{Standard} & \colhead{Factor} \\
\colhead{(1)} & \colhead{(2)} & \colhead{(3)} & \colhead{(4)} &
\colhead{(5)} & \colhead{(6)} & \colhead{(7)} & \colhead{(8)}
}
\startdata
HD 150193 & 2000 May 13\tablenotemark{a} & 05:30 -- 06:10 & 20 & 
1.00 -- 1.01 & $\gamma$ Cru & 1.46 & \nodata \\
HD 100546 & 2000 May 13\tablenotemark{a} & 04:15 -- 05:15 & 32 & 
1.51 -- 1.67 & $\gamma$ Cru & 1.46 & \nodata \\
HD 179218 & 1998 June 23\tablenotemark{b} & 07:03 -- 07:23 &  8 & 
1.15 -- 1.11 & $\alpha$ Her & 1.14 & 1.0 \\
HD 179218 & 1998 June 24\tablenotemark{b} & 07:32 -- 08:10 & 16 & 
1.11 -- 1.12 & $\alpha$ Boo & 1.63 & 0.93 \\
HD 179218 & 1998 June 25\tablenotemark{b} & 08:56 -- 09:16 &  8 & 
1.13 -- 1.16 & $\beta$ Peg & 1.17 & 1.0 \\
HD 179218 & 1998 June 27\tablenotemark{c} & 07:52 -- 08:00 &  4 & 
1.11 -- 1.11 & $\alpha$ Her & 1.18 & 0.92 \\
HD 179218 & 1998 June 30\tablenotemark{c} & 05:24 -- 06:00 & 16 & 
1.29 -- 1.19 & $\alpha$ Her & 1.13 & 1.21 \\

\enddata
\tablenotetext{a}{ \ Wavelength range: $7.4814 - 13.222$~\micron}
\tablenotetext{b}{ \ Wavelength range: $7.5280 - 13.122$~\micron} 
\tablenotetext{c}{ \ Wavelength range: $7.8110 - 13.356$~\micron}

\end{deluxetable}

\begin{deluxetable}{lcc}
\tablewidth{0pc}
\tablecaption{IR PHOTOMETRY OF HD 179218 (2001 April 30.42 UT) 
\label{tab:obsbolo}}
\tablehead{
\colhead{Filter}  & \colhead{Observed} & \colhead{Flux at Zero} \\
\colhead{Wavelength} & \colhead{Magnitude} & \colhead{Magnitude} \\
\colhead{(\micron)} & & \colhead{(W cm$^{-2}$ \micron$^{-1}$)}
}
\startdata
1.25  $[J]$ & $7.19 \pm 0.10$ & $2.90 \times 10^{-13}$ \\
1.65  $[H]$ & $6.71 \pm 0.09$ & $1.12 \times 10^{-13}$ \\
2.34  $[K]$ & $5.88 \pm 0.03$ & $3.39 \times 10^{-14}$ \\
3.65  $[L]$ & $4.73 \pm 0.03$ & $6.43 \times 10^{-15}$ \\
4.90  $[M]$ & $3.99 \pm 0.11$ & $1.99 \times 10^{-15}$ \\
10.00 $[N]$ & $0.88 \pm 0.03$ & $1.23 \times 10^{-16}$ \\
7.91 & $1.71 \pm 0.06$ & $3.08 \times 10^{-16}$ \\
8.81 & $1.24 \pm 0.05$ & $2.02 \times 10^{-16}$ \\
9.80 & $0.81 \pm 0.04$ & $1.33 \times 10^{-16}$ \\
10.27 & $0.65 \pm 0.06$ & $1.11 \times 10^{-16}$ \\
11.70 & $0.47 \pm 0.06$ & $6.63 \times 10^{-17}$ \\
12.49 & $0.60 \pm 0.08$ & $5.13 \times 10^{-17}$ \\
18.30 & $-1.00 \pm 0.09$ & $1.14 \times 10^{-17}$ \\
23.00 & $-1.41 \pm 0.12$ & $4.58 \times 10^{-18}$ \\
\enddata
\end{deluxetable}

\begin{deluxetable}{lccc}
\tablewidth{0pc}
\tablecaption{DATASET SCALING FACTORS TO HIFOGS \label{tab:sedscale}}
\tablehead{
\colhead{Object} & \colhead{{\it ISO} SWS} & \colhead{IRAS} &\colhead{MLOF}\\
\colhead{(1)} & \colhead{(2)} & \colhead{(3)} & \colhead{(4)}
}
\startdata
HD 150193 & 0.90 & 0.62 & \nodata \\
HD 100546 & 0.23 & 0.22 & \nodata \\
HD 179218 & 1.08 & 0.78 & 1.0 \\
\enddata

\end{deluxetable}

\begin{deluxetable}{lll}
\tablewidth{0pc}
\tablecaption{MODEL INPUT PARAMETERS \label{tab:params}}
\tablehead{
 & & \colhead{Observed,} \\
\colhead{Parameter} & \colhead{Definition} & \colhead{Fixed or Varied}  \\
\colhead{(1)} & \colhead{(2)} & \colhead{(3)} 
}
\startdata
$M_{\star} (M_{\odot}) $ & stellar mass  & observed \tablenotemark{a}  \\
$R_{\star} (R_{\odot}) $ & stellar radius  & observed \tablenotemark{a}  \\
$T_{\star}$ (K)          & stellar temperature  & observed \tablenotemark{a} \\
$\Sigma_{\circ} = \Sigma a^{3/2}$ (g cm$^{-2}$) & surface density at 1 AU & 
     fixed at 1000 \\
$r_{\circ}$ (AU) & outer disk radius & varied \\
$H/h$ & visible photospheric height/gas scale height & varied \\
$q$ & slope of grain size distribution interior/surface & varied\\
$r_{max,s}$ (\micron) & maximum grain radius in surface & fixed at 1 \\
$r_{max,i}$ (\micron) & maximum grain radius in interior & fixed at 1000 \\
$T_{sub, Fe}$ (K) & iron sublimation temperature & fixed at 2000 \\
$T_{sub, Ol}$ (K) & olivine sublimation temperature &
        fixed at 1500 \\
$T_{sub, ice}$ (K) & H$_2$O ice sublimation temperature & fixed at 150 \\
\enddata

\tablenotetext{a}{ \ van den Ancker et al.\ (1998)}

\end{deluxetable}

\begin{deluxetable}{ll}
\tablewidth{0pc}
\tablecaption{Minerals \label{tab:mins}}
\tablehead{
\colhead{Dust Type} & \colhead{Reference} \\
\colhead{(1)} & \colhead{(2)} 
}
\startdata
Amorphous Olivine & Dorschner et al.\ (1995) \\
Metallic Iron & Pollack et al.\ (1994) \\
Water Ice & Warren (1984) \\
Mg-pure Crystalline Olivine & J\"ager et al.\ (1998) \\
\enddata

\end{deluxetable}

\begin{deluxetable}{lcccccccccc}
\tablewidth{0pc}
\tablecaption{Stellar Properties and Fitted Parameters \label{tab:fitparams}}
\tablehead{
\colhead{Object} & \colhead{$M_{\star}$} & \colhead{$R_{\star}$} & 
\colhead{$T_{\star}$} & \colhead{$r_{\circ}$}  & \colhead{$H/h$} & 
\colhead{$q$} & \colhead{$f_{i,cr}$ \tablenotemark{a}} & 
\colhead{$f_{o,cr}$ \tablenotemark{b}} &
\colhead{$M_{disk}$ \tablenotemark{c}} &
\colhead{$H(a_{\circ})/a_{\circ}$ \tablenotemark{c}}\\
 & \colhead{$(M_{\odot})$} & \colhead{$(R_{\odot})$} & \colhead{(K)} &
\colhead{(AU)} & & & & & \colhead{$(M_{\odot})$} &  \\
\colhead{(1)} & \colhead{(2)} & \colhead{(3)} & \colhead{(4)} & \colhead{(5)} &
\colhead{(6)} & \colhead{(7)} & \colhead{(8)} & \colhead{(9)} & \colhead{(10)}&
\colhead{(11)}
}
\startdata

HD 150193 & 2.3 & 2.1 & 9333 & 5 & 3 & 3.5 & \nodata & \nodata & 0.003 & 0.11\\
HD 100546 & 2.5 & 1.7 & 10471 & 150 & 4 & 2.5 & 0.43 & 0.27 & 0.024 & 0.42 \\
HD 179218 & 4.3 & 5.2 & 10700 & 150 & 1 & 3.5 & 0.49 & 0.08 & 0.024 & 0.09 \\

\enddata

\tablenotetext{a}{ \ crystalline-to-amorphous silicate ratio for inner disk
($\leq 5$~AU).}
\tablenotetext{b}{ \ crystalline-to-amorphous silicate ratio for outer disk
(5 -- 150~AU).}
\tablenotetext{c}{ \ Disk mass (gas and dust) and maximum aspect ratio of the 
fitted disk are derived, not fitted or input, parameters.}

\end{deluxetable}

\clearpage

\end{document}